\documentclass[useAMS,usenatbib,usegraphicx]{mn2e}

\newcommand{\tev}{TeV}

\newcommand{\chandra}{\textit{Chandra}}

\newcommand{\sgra}{Sgr A*}
\newcommand{\grs}{$\gamma$-rays}
\newcommand{\fe}{Fe~K$\alpha$}
\newcommand{\etal}{et al.}

\title[Time dependence of the TeV source at the GC]
  {Modelling the time dependence of the TeV $\gamma$-ray
  source at the Galactic Centre}
\author[D.\ R.\ Ballantyne, M.\ Schumann \& B.\ Ford]
  {D.~R.~Ballantyne,$^1$\thanks{david.ballantyne@physics.gatech.edu}
  M.~Schumann$^{1,2}$ and B.~Ford$^1$\\
  $^1$Center for Relativistic Astrophysics, School of Physics, Georgia
  Institute of Technology, 837 State Street, Atlanta, GA 30332-0430, USA \\
  $^2$George W. Woodruff School of Mechanical Engineering, Georgia
  Institute of Technology, 801 Ferst Drive, Atlanta, GA 30332-0405, USA}

\date{Accepted 2010 August 15. Received 2010 August 05; in original
  form 2010 July 06}

\pagerange{\pageref{firstpage}--\pageref{lastpage}}
\pubyear{2010}

\usepackage{times}

\begin{document}

\label{firstpage}

\maketitle

\begin{abstract}
The physical mechanism behind the TeV $\gamma$-ray source observed at
the centre of the Galaxy is still unknown. One intriguing possibility
is that the accretion flow onto the central supermassive black hole is
responsible for accelerating protons to TeV energies which then
diffuse outward to interact with molecular gas at distances of $\sim
1$~pc. Here, we build on our earlier detailed calculations of the
proton transport to consider the time and energy dependence of the TeV
signal following a burst of particle acceleration at \sgra. We find
that, due to the strong energy dependence of the proton diffusion, any
variability in the particle acceleration rate will only be visible in the TeV
signal after a delay of $\sim 10$~yrs, and only at energies $\ga 10$~TeV. If the
accelerator is long-lived it must have been running for at least
$10^4$~yrs and have a hard proton injection spectrum of
$\alpha=0.75$ (where $dn/dE_{\mathrm{inj}} \propto E^{-\alpha}$) in
order to produce the correct amount of high energy $\gamma$-ray
flux. This rapid diffusion of high energy protons also rules out the
possibility that the observed TeV source is directly related to the
period of increased activity of \sgra\ that ended $\sim 100$~yrs
ago. However, a good fit to the observed H.E.S.S. data was found with
$\alpha=2.7$ for the scenario of a brief ($\sim$few year long) burst of
particle acceleration that occurred $\sim 10$~yrs ago. If such bursts
are common then they will keep the TeV source energised and will
likely produce spectral variability at $\ga 10$~TeV on $\la 5$~yr timescales. This
model also implies that particle acceleration may be an important
mechanism in reducing the radiative efficiency of weakly accreting
black holes.
\end{abstract}


\section{Introduction}
\label{sect:intro}
The centre of the Milky Way galaxy is located at the position of the
radio source \sgra, and contains a supermassive black hole with a mass
now measured to be $4.5\times 10^6$~M$_{\odot}$ \citep{ghez08}. The proximity of the Galactic centre --- at a distance of only 8.4
kpc \citep{ghez08} --- allows an unprecedented opportunity to
study in detail the complex physics that occurs in the environments
around supermassive black holes. The recent discovery of TeV \grs\
emanating from a region only $\sim 3$~pc from \sgra\
\citep{aha04,kos04,tsu04,alb06,aha06} is an excellent example of the numerous
physical puzzles presented by the Galactic centre. Although the centroid of the Galactic centre TeV source is located
only $\sim 8^{\prime\prime}$ from \sgra\ \citep{ace10}, the origin of the
TeV photons is not understood. There are two main pathways to produce
TeV \grs: the first is a hadronic process where protons with TeV
energies scatter off lower energy protons producing pions ($pp
\rightarrow pp\pi^0\pi^\pm$), with the $\pi^0$s quickly decaying into
two $\gamma$-ray photons. The energetics of the collisions generally produce \grs\
with energies $\sim 10\times$ less than the incident TeV proton. The
second mechanism is via inverse Compton scattering of UV photons off
of TeV electrons. Both mechanisms require a powerful particle accelerator to produce the high-energy particle
population. Unfortunately, the region around \sgra\ contains several
such accelerators, including a supernova remnant \citep{crock05}, a pulsar
wind nebula \citep{wlg06,ha07}, and \sgra\ itself
\citep{ad04,an05a,an05b,ball07}. The centroid of the TeV source has
now been measured with sufficient accuracy to exclude the Sgr A East
supernova remnant as the source of the emission \citep{ace10}; however,
the pulsar and the black hole remain plausible sources of the
high-energy particles responsible for the TeV emission.

The temporal relationship between particle acceleration and TeV
emission is one way to discriminate between the various origins of the
high-energy particles. \sgra\ has recently exhibited flares on timescales
as short as minutes at both X-ray and near-IR wavelengths \citep[e.g.,][]{por03,eck06}. This timescale indicates the flaring
region must be extremely close to the black hole event horizon, one of
the possible sites of particle acceleration \citep{liu06}. Thus, the expectation is if \sgra\ is the origin of the TeV
particles, the TeV \grs\ should be variable, and possibly be
coincident with the flares observed at lower energies. However, no variability of
the TeV source has ever been observed on timescales ranging from
30 minutes to years \citep{rh05,alb06,aha08}. Moreover, during a
coordinated monitoring campaign with \chandra, the H.E.S.S.\
instrument did not detect any increase in TeV $\gamma$-ray emission
during a X-ray flare \citep{aha08}. Thus, the \grs\ are unlikely to be
produced in the same region as the relativistic X-ray emitting
electrons (i.e., within $\sim 100$~Schwarzschild radii of the black
hole). This fact must be explained by any model for the \tev\ \grs,
and seems to support the scenario where the \grs\ are associated with
electrons accelerated by the pulsar wind nebula.

However, protons may be accelerated close to the black hole, but be
converted to \grs\ only after travelling a significant distance away
from the acceleration region \citep[e.g.,][]{ad04,an05b,ball07}. In
the scenario presented by \citet{ball07} proton acceleration was
assumed to occur at distances of only $\sim 20$--$30$ Schwarzschild radii from the black hole \citep[e.g.,][]{liu06}. The particles
would then diffuse away from \sgra\ through the magnetised turbulent ISM,
until possibly colliding with the dense molecular gas in the circumnuclear
disk (CND; e.g., \citealt{mc09}) at a distance of $1$--$2$~pc. \citet{ball07}
modelled this process in detail, paying close attention to the
propagation of the high-energy protons, and obtained a good fit to the observed
TeV spectrum. A key finding of this calculation was the strong energy dependence of
the proton interaction with the CND (the site of the
$\gamma$-ray production). Only $\sim 5$\% of the 100~TeV protons interacted
with the CND to produce \grs, with this rising to $\sim 40$\% at
10~TeV and $\sim 70$\% at 1 TeV. The higher energy protons scattered
far fewer times in the ISM than the low energy ones and therefore
were less likely to encounter the CND and produce
\grs. Thus, a high
energy roll-over in the TeV spectrum at energies $\ga 10$~TeV is a generic
prediction of the hadronic model, and a relatively flat proton
spectrum is required to 
compensate for the low interaction rate between the high energy protons
and the CND. \citet{ball07} only considered the steady-state
$\gamma$-ray flux and spectrum, but it is clear that the significant
energy and geometric dependence of the diffusion and scattering
processes will lead to an interesting time dependence of the
$\gamma$-ray flux if there is some change at the acceleration site. It
is clearly important to quantify the spectral and time dependence of
the expected TeV signal to allow testing of the hadronic model with
the continued $\gamma$-ray monitoring of \sgra. In addition,
 the time evolution of the $\gamma$-ray spectrum may be relevant to the recent \fe\ line maps of the
Galactic centre environment which indicate that \sgra\ was more luminous $\sim
100$~years ago \citep[e.g.,][]{muno07,inui09,ponti10,terr10}. These
observations raises the possibility that the current observed TeV
emission is related to an earlier, more active state of the black hole
and not its current state. 

This paper makes use of the proton transport calculations
 performed by \citet{ball07} to investigate the time and energy
 dependence of the TeV $\gamma$-ray emission in the scenario where the
 protons are accelerated very close to the black hole. We will
 quantify the variability timescale expected for the TeV emission in
 this model and explore the possibility that the current TeV flux
 is related to an earlier outburst of \sgra. The next section details how the time
 dependence of the TeV flux was computed. Section~\ref{sect:res}
 presents the results, while Section~\ref{sect:discuss} contains our
 discussion and conclusions.

\section{Calculations}
\label{sect:calc} 
\citet{ball07} calculated over 220,000 proton trajectories in a
6~pc$\times $6~pc$\times$6~pc cube consisting of $10^6$ equally spaced
cells centred on the Galactic centre. Details of the computations of
the proton transport and $\gamma$-ray production can be found in that
paper. The \grs\ are produced from clumps of high density ($n_{\mathrm{H}} >
3000$~cm$^{-3}$) gas in a model CND that has an inner radius of 1.2~pc
from \sgra\ and a thickness of 1~pc. The spectrum of protons that escapes the
acceleration region at \sgra\ is a power-law with index $\alpha$ (i.e., the
injected proton spectrum is $dn/dE_{\mathrm{inj}} \propto
(E/E_{\mathrm{min}})^{-\alpha}$, where
$E_{\mathrm{min}}=1$~TeV). The steady-state calculation of
\citet{ball07} found a good fit to
the H.E.S.S. TeV data with $\alpha=0.75$, a much flatter spectrum
than is expected from traditional particle acceleration
mechanisms (where $\alpha \sim 2$--$2.5$). We compute $\gamma$-ray
flux (in ph~m$^{-2}$~s$^{-1}$~TeV$^{-1}$) at 31 logarithmically spaced
energies: $\log (E/\mathrm{eV})=11,11.1,\ldots,13.9,14$.

The time step used in the proton transport computations varied between a
small fraction of a year for the lowest energy particles to just over a
year for protons with energies close to 100~TeV. Thus, variations in the
TeV spectrum and flux can only be calculated for timescales greater
than $\sim 1$~yr. The time dependence
of the $\gamma$-ray flux will be investigated in two ways. First, the
particle accelerator located at the origin
is assumed to turn on at $t=0$ and remain on at a constant rate. All proton trajectories are followed for an
observed time $t$, and the optical depth to scattering with CND gas during this time is recorded. After a complete
census of the optical depth to proton-proton scattering in the dense
gas is determined, the total $\gamma$-ray flux in multiple energy
bands is then calculated. The limiting
time $t$ is increased until the fluxes reach the observed values
determined by the steady-state model.

A second calculation is performed to investigate the response of the
TeV flux to a burst of particle acceleration lasting $\Delta t$
years. The proton trajectories are followed and the $\gamma$-ray flux
calculated in the time interval ($t$,$\Delta t+t$) with $t$ increasing
from $0$ to $105$~yrs in steps of $5$~yrs. This computation is performed for
bursts of activity lasting $\Delta t=3$, $10$, $30$, $100$ and
$300$~yrs. This experiment will help determine if the current observed
TeV source may have been caused by an earlier period of activity
around \sgra. 

\section{Results}
\label{sect:res}
\subsection{Rise Time to a Constant Flux}
\begin{figure}
\centerline{
\includegraphics[width=0.35\textwidth,angle=-90]{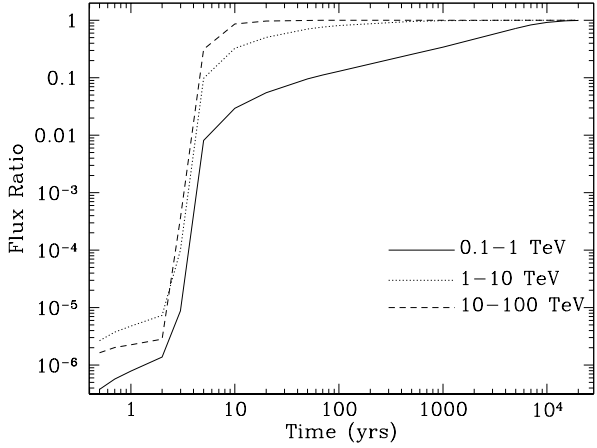}
}
\caption{The time evolution of the Galactic centre TeV source in
  three different energy bands. Proton acceleration is assumed to
  begin at the origin (i.e., very close to \sgra) at $t=0$ and
  continue at a constant rate. The spectrum of protons injected into
  the interstellar medium is a power-law with spectral index
  $\alpha=0.75$. The fluxes are plotted as a fraction of the observed
  H.E.S.S. values. Due to the rapid diffusion time of the very high
  energy protons, the $10$--$100$~TeV flux reaches $>90$\% of its steady-state
  value after only $\sim 10$~yrs. In contrast, it takes several
  hundred years for the $1$--$10$~TeV flux and $\sim 10^4$~yrs for the
  $0.1$--$1$~TeV flux to reach the same fraction of the steady-state
  value. The large increase in the flux ratios after $\sim 5$~yrs
  corresponds to the proton `front' reaching the CND
  surrounding \sgra.}
\label{fig:fratio1}
\end{figure}

Figure~\ref{fig:fratio1} plots the $\gamma$-ray flux as a function of time in three
different TeV energy bands.
This figure shows the fluxes as fractions of the observed
H.E.S.S. values as determined by the steady-state model \citep{ball07}. The Lorentz force of the magnetic field on a proton is proportional to
$1/\gamma$, where $\gamma$ is the Lorentz factor of the proton, so high energy protons spend less time random walking
around the interstellar medium and reach the dense gas much more
rapidly than the lower energy particles. Therefore, it only takes
$\sim 10$~yrs for the $10$--$100$~TeV flux to reach a significant
fraction of its steady-state value. The lower energy protons take much
longer to diffuse outwards. Indeed, after the particle accelerator has
been operating for $10$~yrs the $0.1$--$1$~TeV flux has reached only
$\sim 3$\% of its final value. The accelerator must run for over
$10^4$ years\footnote{This is $\sim 1000\times$ shorter than the
  estimated lifetime of the CND \citep{mc09}; thus, these timescales
  are not effected by the static gas model for the CND and surrounding
  ISM.} for enough protons to interact with the molecular gas to
account for the observed H.E.S.S. flux. The normalisation of the
injected proton spectrum was chosen so that these equilibrium fluxes
would match the observed values. If a higher normalisation was chosen
(i.e., the proton luminosity produced by the \sgra\ accelerator was
higher) then the timescales for the fluxes to match the observed
values would be shorter, but the steady-state fluxes would be greater
than what is observed, violating the initial assumption of our
scenario (that the observed TeV source is due to an accelerator has
been operating at a constant level for a long time). 

\begin{figure}
\centerline{
\includegraphics[width=0.35\textwidth,angle=-90]{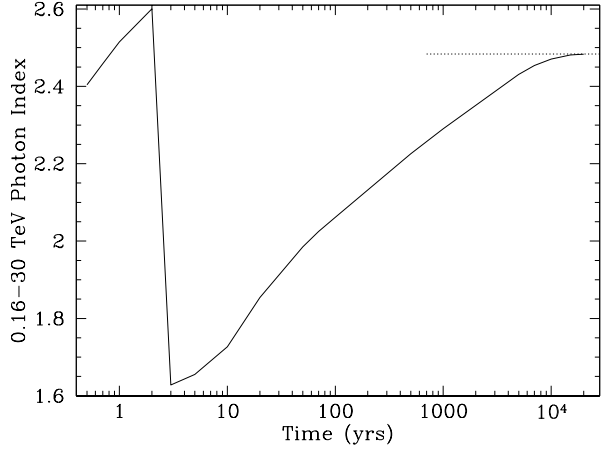}
}
\caption{The time evolution of the $0.16$--$30$~TeV photon index for
  the scenario where a constant proton accelerator is ignited at the origin at
  $t=0$. The dotted line corresponds to a photon index of $2.48$, the
  value derived from the best fit model to the H.E.S.S. data. This
  value is somewhat steeper than the observed photon index of $2.25
  \pm 0.10$ \citep{aha06} because of the roll over in high energy
  \grs\ predicted by the model. As the low energy protons diffuse much slower than
  the higher energy ones, the TeV spectrum softens slowly with time,
  and does not reach the observed value for over $10^4$~yrs. The sharp
  increase and decrease of the photon index after only 2 years is due
  to the interaction of a small number of primarily low-energy protons with one dense clump of gas that is
  located only $\sim 0.1$~pc away from the origin.}
\label{fig:slopevstime}
\end{figure}

After the accelerator has
been operating for about 3~yrs, the first protons start interacting
with the CND and produce TeV \grs. As seen in
Figure~\ref{fig:slopevstime}, the TeV spectrum produced at this time
would be hard with a $0.16$--$30$~TeV photon index $\Gamma \sim 1.6$ (where photon flux
$\propto E^{-\Gamma}$), as only a small fraction of the low energy
protons would be on fortuitous trajectories to interact with the high
density gas at this early stage. 
The spectrum softens steadily over the next $10^4$ years as the lower
energy protons reach the CND and produce \grs.

Several conclusions can be drawn from this experiment. First, if there
is a proton accelerator close to \sgra\ and it either switches on or
increases its acceleration rate (i.e., a flare of protons), the
resulting \grs\ will not be detected for $\sim 10$~yrs and only in the
very high energy $10$--$100$~TeV band. It will take over a century for
the flare to be seen in $1$--$10$~TeV \grs. Therefore, the null result
found in the coordinated X-ray/TeV monitoring of \sgra\ \citep{aha08}
indicates that the X-ray and TeV photons are being produced in
different regions, but does not rule out that \sgra\ is ultimately
responsible for the proton acceleration. These results show that time
variability of the TeV source is likely to be observed only in the
$10$--$100$~TeV band and on decade-like timescales. If such
variability was observed, however, it would give strong support to a
high-energy diffusion process producing the \grs\ and thus the
hadronic and \sgra\ model for the TeV emission. On the other hand, if
no variability of the TeV source is ever detected, it still will not
rule out \sgra\ as the site of particle acceleration, but it will
imply that this accelerator must be operating at a relatively constant
rate for $>10^4$ yrs.

\subsection{Is the TeV Source Connected to Past Activity of
  \sgra?}
\begin{figure}
\centerline{
\includegraphics[width=0.35\textwidth,angle=-90]{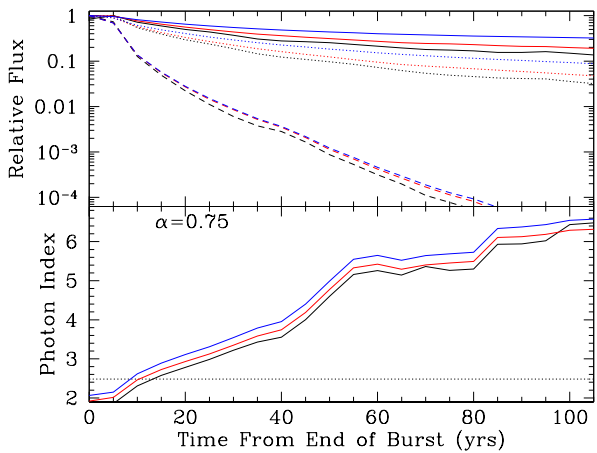}
}
\caption{The time evolution of the TeV $\gamma$-ray source following
  bursts of proton injection that lasted 10 (black lines), 30 (red
  lines) and 100 (blue lines) year. The top panel plots how the
  $0.1$--$1$ (solid line), $1$--$10$ (dotted line) and $10$--$100$
  (dashed line)~TeV fluxes evolve with time, with each band normalised to
  its maximum flux. The bottom panel shows the time evolution of the $0.16$--$30$~TeV photon index. The dotted line corresponds to a photon index of $2.48$, the
  value derived from the best fit model to the H.E.S.S. data. These
  results were computed assuming $\alpha=0.75$.}
\label{fig:twoplot}
\end{figure}

In the hadronic model for the TeV source, the \grs\ are produced from
protons that have diffused outwards from the acceleration site close
to \sgra. The previous section showed that this diffusion can cause a
significant delay between when the protons were accelerated and when
they produce \grs. Given the evidence that points to a more luminous black hole at the Galactic
centre $\ga 100$~yrs ago, it is interesting to consider the
possibility that the current observed $\gamma$-ray source was caused
by this earlier activity. Unfortunately, Figure~\ref{fig:twoplot}
rules out this possibility.
The rapid diffusion of the $>10$~TeV protons causes the resulting
$\gamma$-ray spectrum to soften rapidly. As discussed above, the highest
energy protons scatter off the CND and
produce high energy \grs\ only $\sim 10$~yrs after being injected into
the interstellar gas. Increasing the duration of the burst of particle
acceleration does not improve this situation as long as the
observations are being made following the end of the burst. This is
because it is only the protons that were injected within the last
decade that produce the high energy $\ga 10$~TeV
\grs. Fig.~\ref{fig:twoplot} shows the results for an injected proton
spectrum with $\alpha=0.75$. The same calculations were also performed
with more traditional values of $\alpha=2$ and $2.5$, but, as the
diffusion properties of the protons are independent of $\alpha$, the
conclusion is unchanged: if the observed TeV signal is related to
accretion onto \sgra, it cannot be caused by a burst of protons
produced $\ga 100$~years ago during the period of greater X-ray luminosity.

\begin{figure}
\centerline{
\includegraphics[width=0.35\textwidth,angle=-90]{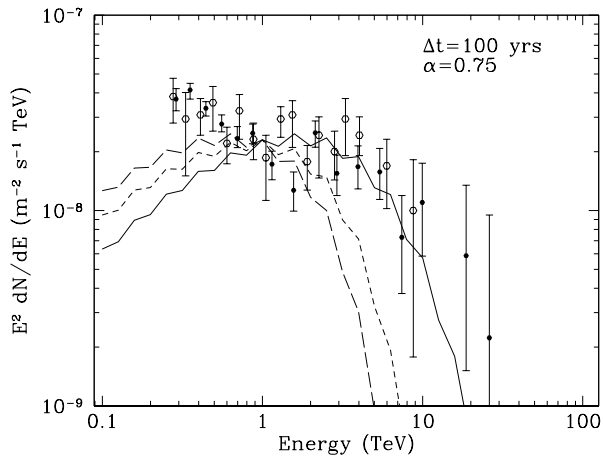}
}
\caption{The TeV spectrum produced at 10 (solid line), 50
  (short-dashed line) and 100 (long dashed-line) years after a
  100 year-long burst of proton acceleration close to \sgra. The 2003
  H.E.S.S. data are plotted as the open points, and the 2004
  H.E.S.S. data are plotted as the solid points
  \citep{aha04,aha06}. All three spectra have been normalised to a
  flux of $2.3\times 10^{-8}$~m$^{-2}$~s$^{-1}$~TeV at 1~TeV.}
\label{fig:spect1}
\end{figure}

Figure~\ref{fig:twoplot} indicates that the TeV spectrum may have the
right shape $\sim 10$--$15$~years after a burst; however, comparing
these spectra to the H.E.S.S. data illustrates another feature of the proton
diffusion in the Galactic centre. Figure~\ref{fig:spect1} plots the
TeV spectrum produced at 10, 50 and 100~yrs following a 100 year-long
burst of proton acceleration at the Galactic centre.
This figure shows that at energies less than 1~TeV, all three spectra
retain the hard shape of the injected proton spectrum, in marked
contrast to the observed spectrum at those energies. Thus, even
though the average spectral slope 10 years after the burst is close to
the observed value, the actual shape of the spectrum does not describe
the data. The reason for this very hard spectrum at $\la 1$~TeV is
that the low energy protons take a very long time to diffuse outward
to the CND (see Fig.~\ref{fig:fratio1}). The low
energy \grs\ that make up the plotted spectra are made up from very
few interactions of protons and dense clouds. As time passes, more of
these interactions occur and, since a very high percentage of the low energy protons interact
 with the molecular gas, the $0.1$--$1$~TeV spectrum
 rises steadily. However, as seen in Fig.~\ref{fig:fratio1}, it will
 take several thousand years before the majority of injected low
 energy protons have interacted with the CND. Therefore,
 the burst scenario with $\alpha=0.75$ leaves us with a catch-22
 situation: the high energy $\gamma$-ray spectrum can only be matched
 $\sim 10$~yrs after the burst, but the low energy TeV spectrum will
 not have the right shape for several thousand years after the burst. 

\begin{figure}
\centerline{
\includegraphics[width=0.35\textwidth,angle=-90]{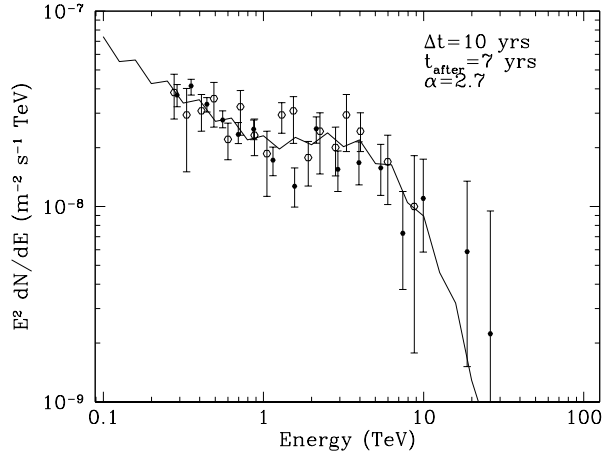}
}
\caption{The TeV spectrum produced 7 years after the end of a
  10 year-long burst of proton acceleration close to \sgra. The
  injected proton spectrum has a spectral index of $2.7$. The spectrum has been normalised to a
  flux of $2.3\times 10^{-8}$~m$^{-2}$~s$^{-1}$~TeV at 1~TeV. The 2003
  H.E.S.S. data are plotted as the open points, and the 2004
  H.E.S.S. data are plotted as the solid points \citep{aha04,aha06}.}
\label{fig:spect2}
\end{figure}

These results lead to an interesting possibility: as the low energy TeV spectrum closely follows the injection
spectrum after a burst of protons, is it possible to fit all of the H.E.S.S. data with a softer proton
injection spectrum? This would only work for burst durations $\la
10$~yrs so that high energy \grs\ would make up a significant fraction
of all the proton interactions. We searched for such a solution using
$\chi^2$ fitting to the observed H.E.S.S. data. All model spectra were
normalised to a flux of $2.3\times 10^{-8}$~m$^{-2}$~s$^{-1}$~TeV at
1~TeV with $\alpha$ and the time since the end of the burst being free
parameters. With $\Delta t=10$~yrs we found a good fit
($\chi^2/$d.o.f.$=43.4/31$, where d.o.f.$=$ degrees of freedom) at
$7$~yrs after the burst and $\alpha=2.7$. The resulting spectrum is
shown in Figure~\ref{fig:spect2}. A similar fit
($\chi^2/$d.o.f.$=42.3/31$) was found for $\Delta t=3$~yrs with
$\alpha=2.7$, but now observed 10~yrs after the end of the burst.
These spectra are not as good a fit to the data as the steady-state
model, but they have the advantage of a softer
proton spectrum that is more consistent with particle acceleration
processes \citep[e.g.,][]{liu06,rey08}. The observed $\gamma$-ray flux and the
calculated energy-dependent proton interaction rate allows the
calculation of the $1$--$40$~TeV energy in protons produced by \sgra\
in this model. We find this energy to be $2\times 10^{47}$~ergs and
$5\times 10^{47}$~ergs for the $\Delta t=10$ and $3$~yrs fit,
respectively. Assuming a constant proton production rate during the
burst, this results in proton luminosities of $5\times
10^{38}$~erg~s$^{-1}$ and $5\times 10^{39}$~erg~s$^{-1}$,
respectively. We will consider the viability of these fits in the next
section.

\section{Discussion and Conclusions}
\label{sect:discuss}
This paper has considered the time dependence of the Galactic centre
TeV source under the scenario that the \grs\ originate from
protons that were accelerated close to the black hole located at \sgra. First,
we computed the time evolution of the TeV source following the
ignition of a long-lived accelerator. This experiment will also be
applicable to the situation where a flare or a burst of proton
acceleration occurred at some point during the accelerator's
lifetime. We find that, as a result of the energy dependent diffusion
 rate of the protons, the shortest timescale for
variability is $\sim 10$~yrs and occurs only for $\ga 10$~TeV \grs.
Detection of this energy dependent variability would be compelling
evidence for the hadronic model and an acceleration site close to
\sgra. If the TeV source is a result of a long-lived accelerator than
it must have been operating for at least $10^4$~yrs to build up the
large $0.1$--$1$~TeV flux. However, because of the strong energy
dependent interaction rate of the protons with the molecular gas, the
injected proton spectrum must be very hard with $\alpha=0.75$.

We also considered the possibility that the accelerator was not
long-lived and that the observed \grs\ are a result of a recent burst
of particle acceleration that has since ended. This model seemed
particularly interesting in light of the recent results from \fe\
reverberation indicating that \sgra\ was more luminous $\ga 100$~yrs
ago \citep[e.g.,][]{ponti10,terr10}. However, we are able to definitively rule out this possibility,
as there is no way that a burst of protons of any duration or spectral
index can reproduce the observed photon index over 100~yrs after the
end of the burst. This is because the highest energy protons produce
their \grs\ within $10$~yrs of being released into the interstellar
medium. Moreover, if $\alpha=0.75$ it would take several thousand
years after the end of the burst to reproduce the shape of the low
energy H.E.S.S. spectrum. 

Considerations of how the TeV spectrum evolved with time allowed the
discovery of new solution to the observed H.E.S.S. spectrum, now with
$\alpha=2.7$. This new model implies the H.E.S.S. observations were
made $\sim 10$~yrs after a burst lasting $3$--$10$~yrs. This model has
the immediate advantage that the proton spectral index is easily
accounted for by particle acceleration processes, but implies a more
random and variable acceleration region. This would not be unexpected
given the observed variability (on much shorter timescales) seen from
\sgra\ in X-ray wavelengths. It should be emphasised that burst
durations shorter than 3~yrs and (slightly) longer than 10~yrs would
still be able to fit the H.E.S.S. data. The only differences from the
above result would be in the delay between the end of the burst and
the observation time, and in the proton luminosity produced by the
accelerator (a higher luminosity would be needed for a shorter $\Delta
t$). A key prediction of this scenario is that the spectrum will soften
noticeably over the next $\sim 5$~yrs as the high energy \grs\
disappear. Of course, further short bursts of protons could continue
to energise this region of the spectrum. In that case, some spectral
variability in the high-energy region of the spectrum might be
observed on $\sim$few year long timescales as the spectral index of
the protons would be unlikely to be identical during each outburst.

The fits for the burst scenario imply a proton
luminosity of $\sim 10^{38-39}$~erg~s$^{-1}$. This is
within an order of magnitude of the proton luminosity originally
calculated by \citet{liu06} in their first application of stochastic
acceleration of protons to the TeV source. It is interesting to
compare this power in protons to the radiative output of \sgra. The
observed bolometric luminosity of \sgra\ is $\sim 10^{36}$~erg~s$^{-1}$,
resulting in an Eddington ratio of $3\times 10^{-9}$. The
gas density and temperature has been measured at the Bondi radius \citep{bag03}, and
the resulting Bondi accretion rate is $\sim
10^{-5}$~M$_{\odot}$~yr$^{-1}$ \citep[e.g.,][]{qua04}. To account for the observed
luminosity of \sgra\ this flow must have an extremely small radiative
inefficiency of $\sim 5\times 10^{-6}$. Accretion models have only
been able to explain this very low efficiency by reducing
the actual accretion rate onto the black hole by $\sim 10^{-3}$ via strong outflows
(\citealt{yqn03}; see also \citealt{shar07}). These low accretion
rates seem to be consistent with those derived from sub-millimetre
polarisation measurements \citep[e.g.,][]{marr07}, although care is
needed in the interpretation of those data \citep[e.g.,][]{bop07}. If
the TeV source is generated by brief bursts of particle acceleration
in the accretion flow, then this could be another mechanism of
reducing the radiative efficiency of weakly accreting black
holes. There already exists compelling
evidence that at very low inflow rates the liberated accretion energy
can be efficiently converted into particle acceleration. For example, the
presence of jets is much more common in low-luminosity AGNs than in
other, more luminous accreting black holes
\citep[e.g.,][]{ho08}. Similarly, accreting stellar mass black holes
are observed to 
preferentially produce jets at low Eddington ratios \citep[e.g.][]{fend01,macc03,gallo06}. At even smaller
accretion rates (but very high black hole mass), there seems to be an efficient conversion of
accretion power into jet power \citep{allen06}. The radio and
sub-millimetre emission from \sgra\ can also be described as
originating from a mildly relativistic jet \citep[e.g.,][]{falcke09,maitra09}. The TeV source from
the Galactic centre could be another indication that weakly accreting
black holes are important sites of particle acceleration,
where, in this case, the efficiency of conversion into proton power
would be $\sim 10^{-3}-10^{-4}\dot{M}_{\mathrm{Bondi}}$. 

In our exploration of the time-dependence of the $\gamma$-ray signal
from the Galactic centre, we have found two potential explanations for
the source. The first requires
a long-lived particle accelerator, lasting at least $10^4$~yrs, and a
very hard proton injection spectrum of $\alpha=0.75$. The
$10^4$~yrs timescale is the minimum required for protons to diffuse
throughout the Galactic ridge and contribute to the observed TeV glow
of the inner galaxy \citep{aha06b}. In addition, \citet{crock07} found that the
interaction of the $\alpha=0.75$ proton spectrum with the stellar wind
gas in the Galactic centre naturally accounted for the observed GHz radio emission from that region. The proton luminosity required
from this particle accelerator would be $\la 10^{34}$~erg~s$^{-1}$ and
would indicate that particle acceleration would not be an important
energetic component of the accretion flow, perhaps consistent with the
small fraction of non-thermal electrons needed to fit the \sgra\
spectrum \citep[e.g.,][]{yqn03}.

The rapid burst scenario reproduces the observed spectrum with a more
typical proton spectrum of $\alpha=2.7$, but requires the
H.E.S.S. observations to have been made at specific times after the
bursts. If the bursts of particle acceleration are reasonably
frequent, however, then this problem is not too constraining. Perhaps
the cleanest discrimination between the two scenarios is that a much
higher degree of spectral variability would be expected in high energy
\grs\ in the burst model than for the long-lived accelerator. In the
former case, the high energy spectrum should soften dramatically over
the next few years unless additional bursts have re-energised the high
energy particles. The long-lived accelerator would be expected to only
show moderate variability in response to changes in the acceleration
rate and, at high energies, this would be at decade-like
timescales. If any spectral variability of the TeV spectrum is
observed, then this would be strong support for the hadronic
model. To conclude, we recommend that spectral monitoring of the
Galactic Centre TeV source should be performed on an annual basis to
search for spectral variability, with particular emphasis in
searching for changes in the $>10$~TeV spectrum. Improving the quality
of the observations at $>10$~TeV will be extremely important in
determining the mechanism behind the TeV source.



\bsp 

\label{lastpage}

\end{document}